\documentclass{article}
\usepackage{spconf,amsmath,epsfig}
\usepackage{amsmath,amssymb,amsthm}%
\newcounter{MYtempeqncnt}
\usepackage{cite}
\usepackage{amsfonts}
\usepackage{fancyhdr}
\usepackage{graphicx}%
\usepackage{listings}%
\usepackage{arydshln}
\usepackage{subfig}%
\usepackage{algorithmic}%
\title{Joint localization and clock synchronization for \\ wireless sensor networks}
\name{Sundeep Prabhakar Chepuri, Geert Leus, and Alle-Jan van der Veen}
\address{Faculty of Electrical Engineering, Mathematics and Computer Science\\ Delft University of Technology, 2628 CD Delft, The Netherlands\\ E-mail: s.p.chepuri@tudelft.nl, g.j.t.leus@tudelft.nl, a.j.vanderveen@tudelft.nl.}
\begin{document}
%
\maketitle
\begin{abstract}
A fully-asynchronous network with one target sensor and a few anchors (nodes with known locations) is considered. Localization and synchronization are traditionally treated as two separate problems. In this paper, localization and synchronization is studied under a unified framework. We present a new model in which time-stamps obtained either via two-way communication between the nodes or with a broadcast based protocol can be used in a simple estimator based on least-squares (LS) to jointly estimate the position of the target node as well as all the unknown clock-skews and clock-offsets. The Cram\'er-Rao lower bound (CRLB) is derived for the considered problem and is used as a benchmark to analyze the performance of the proposed estimator.
\end{abstract}
\section{Introduction}
\label{sec:intro}
Localization and clock synchronization are two key components of any self-organizing location-aware wireless sensor network (WSN). A WSN enables distributed information processing tasks like data sampling, information fusion, and other time-based tasks~\cite{largenetwork}. Every node in the network has an autonomous clock. These individual clocks in a WSN drift from each other due to imperfections in the oscillator, aging and other environmental variations. It is essential to calibrate these imperfections from time to time to achieve a network-wide time coherence. A plethora of algorithms for clock synchronization can be found in~\cite{Yikchungwu_survey}. For the data to be meaningful, the location where the data is acquired is often required. Computing the location of the nodes is commonly called {\it localization}, and is a well-studied topic~\cite{Gusta05SPM}.

Even though localization and clock synchronization are tightly coupled, traditionally they are treated as two separate problems. Recently, for an anchorless and a fully asynchronous network, a global least-squares (GLS) estimator based on a two-way time-stamp exchange protocol for joint clock synchronization and ranging has been proposed in~\cite{raj_camsap}. Exploiting the broadcast nature of the wireless medium, an asymmetrical time-stamping and passive listening (ATPL) protocol was proposed in~\cite{ChepuriSPL} for joint clock synchronization and ranging. Subsequently, the estimated pairwise distances can be used as an input to the least-squares (LS) based range-squared method for localization.  
Joint estimation of the position and the clock parameters of a sensor based on the two-way time-stamp exchange protocol has been considered in \cite{ZhengJSL}, where an synchronous network is considered where only the sensor node suffers from clock-skews and clock-offsets and the anchors are assumed to be synchronized. In~\cite{yiyinATR}, localization of the sensor node in a fully-asynchronous network has been studied, where the main focus is on localization but also certain approximations of the clock parameters are required to solve the problem. 

In this paper, we again consider fully-asynchronous network consisting of one sensor and a few anchors, and investigate localization and clock synchronization under a unified framework. We propose a linear data model for joint localization and clock synchronization. The data model is generic in the sense that it can be used with either the two-way ranging protocol or the ATPL protocol. This is used in an estimator based on LS to jointly estimate the position of the sensor node and all the unknown clock-skews and clock-offsets. The Cram\'er-Rao Lower Bound (CRLB) is derived for the considered problem and is used as a benchmark to analyze the performance of the proposed estimators. \\
{{\textsl{\it Notation}:  \\
Upper (lower) bold face letters are used for matrices (column vectors); 
$(\cdot)^T$ denotes transposition; $\odot$ ($\oslash$) refers to element-wise matrix or vector product (division); $(.)^{\odot2}$ denotes the element-wise matrix or vector squaring; 
$\mathrm{bdiag}(.)$ a block diagonal matrix with the matrices in its argument on the main diagonal; $\mathbf{1}_N$ ($\mathbf{0}_N$) denotes the $N \times 1$ vector of ones (zeros);  $\small \mathbf{I}_N$ is an identity matrix of size $N$; $\mathbb{E}(.)$ denotes the expectation operation; $\otimes$ is the Kronecker product. 
 
\begin{figure*}[!htp]
\vspace*{-10mm}
\normalsize 
\setcounter{MYtempeqncnt}{\value{equation}}
\setcounter{equation}{4}
\begin{equation}\label{eq:example}
\footnotesize
\begin{aligned} 
\overbrace{\left[\begin{array}{cc|cc|cc|cc|ccc}{\bf t}_{01} & {\bf 1}_{K} & -{\bf t}_{10} & -{\bf 1}_{K} & {\bf 0}_K & {\bf 0}_K & {\bf 0}_K & {\bf 0}_K & {\bf e}_{10} & {\bf 0}_K & {\bf 0}_K \\{\bf t}_{02} & {\bf 1}_{K} & {\bf 0}_K & {\bf 0}_K & -{\bf t}_{20} & -{\bf 1}_{K} & {\bf 0}_K & {\bf 0}_K & {\bf 0}_K & {\bf e}_{02} & {\bf 0}_K \\{\bf t}_{03} & {\bf 1}_{K} & {\bf 0}_K & {\bf 0}_K & {\bf 0}_K & {\bf 0}_K & -{\bf t}_{30} & -{\bf 1}_{K} & {\bf 0}_K & {\bf 0}_K & {\bf e}_{03}\end{array}\right]}^{\quad \quad {\bf A} \in \mathbb{R}^{ 3K \times 11}} \overbrace{\begin{bmatrix} {\bf c}_0 \\{\bf c}_1 \\ {\bf c}_2 \\  {\bf c}_3 \\ \boldsymbol{\tau}_{0}\end{bmatrix}}^{\, \boldsymbol{\theta} \,\in \,\mathbb{R}^{11 \times 1}} =  \overbrace{\begin{bmatrix} {\bf n}_{01} \\{\bf n}_{02} \\ {\bf n}_{03} \end{bmatrix}}^{\, \bf{n} \,\in \,\mathbb{R}^{9 \times 1}}.
\end{aligned}
\end{equation}
\begin{equation}\label{eq:example2}
\footnotesize
\begin{aligned} 
\overbrace{\left[\begin{array}{cc|cc|cc|ccc}{\bf t}_{01} & {\bf 1}_{K} & -{\bf t}_{10} & -{\bf 1}_{K} & {\bf 0}_K & {\bf 0}_K & {\bf e}_{10} & {\bf 0}_K & {\bf 0}_K \\{\bf t}_{02} & {\bf 1}_{K} & {\bf 0}_K & {\bf 0}_K & -{\bf t}_{20} & -{\bf 1}_{K} & {\bf 0}_K & {\bf e}_{02} & {\bf 0}_K \\{\bf t}_{03} & {\bf 1}_{K} & {\bf 0}_K & {\bf 0}_K & {\bf 0}_K & {\bf 0}_K & {\bf 0}_K & {\bf 0}_K & {\bf e}_{03}\end{array}\right]}^{\quad \quad {\bf A} \in \mathbb{R}^{ 3K \times 9}} \overbrace{\begin{bmatrix} {\bf c}_0 \\{\bf c}_1 \\ {\bf c}_2 \\ \boldsymbol{\tau}_{0}\end{bmatrix}}^{\, \boldsymbol{\theta} \,\in \,\mathbb{R}^{9 \times 1}} = -\overbrace{\left[\begin{array}{cc}{\bf 0}_K & {\bf 0}_K \\{\bf 0}_K & {\bf 0}_K \\-{\bf t}_{30} & -{\bf 1}_{K}\end{array}\right] \bf{c}_3}^{{\bf t} \in \mathbb{R}^{3K \times 1}} \quad + \quad \overbrace{\begin{bmatrix} {\bf n}_{01} \\{\bf n}_{02} \\ {\bf n}_{03} \end{bmatrix}}^{\, \bf{n} \,\in \,\mathbb{R}^{9 \times 1}}.
\end{aligned}
\end{equation}
\setcounter{equation}{0}
\hrulefill
\end{figure*}

\section{Network model}
\label{sec:systemmodel}

\begin{figure} [!h]
\centering
\includegraphics[width=2in]{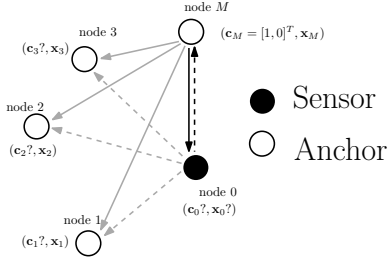}
\caption[An illustration of the network model, together with the know and unknown parameters. Light shaded lines refer passive listening links.] {An illustration of the network model, together with the know and unknown parameters. Light shaded lines refer to the passive listening links.}
\label{fig:asymmetrical}
\vspace*{-4mm}
\end{figure} 
We consider a fully-asynchronous network with $M$ anchors and one sensor ({\it node 0}) as shown in Fig.~\ref{fig:asymmetrical}. We assume one of the nodes has a relatively stable clock oscillator and is used as a clock reference. All the other nodes suffer from clock-skews and clock-offsets.  The network model considered here is the same as the model considered in~\cite{ChepuriSPL}.

All the nodes are distributed over an  $l$-dimensional space, with $l=2$ (plane) or $l=3$ (3-D space). Let the vector ${\bf x}_i \in \mathbb{R}^{l \times 1}$ denote the coordinates of the $i$th node. The coordinates of all the anchors are collected in a matrix ${\bf X} = [{\bf x}_1,{\bf x}_2,\ldots,{\bf x}_M] \in \mathbb{R}^{l \times M}$. The unknown coordinates of the sensor are collected in ${\bf x}_0$. The distance between the $i$th node and the $j$th node is denoted by 
\begin{equation}\label{eq:distance}
\vspace*{-1mm}
\begin{aligned}
d_{ij} = d_{ji} &= \|\mathbf{x}_{i}-\mathbf{x}_{j}\|_2 
= \sqrt{\|\mathbf{x}_{i}\|^2-2\mathbf{x}_{i}^T\mathbf{x}_{j}+\|\mathbf{x}_{j}\|^2}
\end{aligned}
\end{equation}

Let $t_{i}$ be the local time at the $i$th node and $t$ be the reference time. We then assume that the relation between the local time and the reference time can be given by a first-order affine clock model~\cite{raj_camsap},
\begin{equation}\label{eq:clockmodel}
\vspace*{-1mm}
\begin{aligned}
t_{i} = \omega_{i} t + \phi_{i} \quad \Leftrightarrow \quad t = \alpha_{i} t_{i} + \beta_{i}
\end{aligned}
\vspace*{-1mm}
\end{equation}
where $\omega_{i} \in \mathbb{R}_{+}$  is the clock-skew, $\phi_{i} \in \mathbb{R}$ is the clock-offset,  $\alpha_{i}=\omega_{i}^{-1}$ and $\beta_{i} =  -\omega_{i}^{-1}\phi_{i}$ are the synchronization parameters of the $i$th node. Without loss of generality, we use anchor $M$ as absolute time reference, i.e., $[\omega_M,\phi_M]=[1,0]$. The unknown synchronization parameters are collected in $\boldsymbol{\alpha}=[\alpha_0,\alpha_1,\ldots,\alpha_{M-1}]^T$ and $\boldsymbol{\beta}=[\beta_0,\beta_1,\ldots,\beta_{M-1}]^T$. The unknown clock-skews and clock-offsets are respectively given by 
\vspace*{-1mm}
\begin{equation}
\boldsymbol{\omega} = \mathbf{1}_M \oslash \boldsymbol{\alpha}\quad \text{and} \quad \boldsymbol{\phi}  =  - \boldsymbol{\beta} \oslash \boldsymbol{\alpha} \label{eq:omega}.
\end{equation}  

Nodes in the network can communicate with each other either via a two-way communication protocol~\cite{raj_camsap} or an ATPL protocol~\cite{ChepuriSPL} as illustrated in Fig.~\ref{fig:asymmetrical}. \vspace*{-3mm}
\section{Problem formulation}

In this paper, we focus on the two-way communication protocol for deriving the data model. The transmission and reception time-stamps are recorded independently at local time coordinates both during the forward and the reverse links. The time-stamp recorded at the $i$th node when the $k$th iteration message departs to the $j$th node is denoted by $T_{ij}^{(k)}$, and upon arrival of the corresponding message, the $j$th node records the time-stamp $T_{ji}^{(k)}$. For the sake of generality, we do not put any constraints on the sequence of forward links and reverse links or the number of time-stamps recorded~\cite{raj_camsap,ChepuriSPL}. 

The time-of-flight for a line-of-sight (LOS) transmission between the $i$th node and the $j$th node can be defined as $\tau_{ij}  = {\nu}^{-1}{d_{ij}}$, where $\nu$ denotes the speed of a wave in a medium. Using (\ref{eq:clockmodel}), $\tau_{ij}$ can be written in terms of the local clock coordinates as
\begin{equation}\label{eq:timemarkers}
\begin{aligned}
\tau_{ij} &= (\alpha_jT_{ji}^{(k)} + \beta_j) - (\alpha_iT_{ij}^{(k)} + \beta_i) + n_{ij}^{(k)} 
\end{aligned}
\end{equation}
where $n_{ij}^{(k)}$ denotes the aggregate measurement error on the time-stamps. 
\setcounter{equation}{6}

In all there are $K$ time-stamps recorded at each node and the time-stamps recorded at the $i$th node are collected in ${\bf t}_{ij} = [T_{ij}^{(1)},T_{ij}^{(2)},\ldots,T_{ij}^{(K)}]^T \in \mathbb{R}^{K \times 1}$. The direction (forward or reverse) of the $k$th link is denoted by $e_{ij}^{(k)}$, where $e_{ij}^{(k)} = 1$ for transmission from node $i$ to node $j$ and  $e_{ij}^{(k)} = -1$ for transmission from node $j$ to node $i$.  The direction information is collected in a vector ${\bf e}_{ij} = [e_{ij}^{(1)},e_{ij}^{(2)},\ldots,e_{ij}^{(K)}]^T \in \mathbb{R}^{K \times 1}$. The error vector is denoted by ${\bf n}_{ij} = [n_{ij}^{(1)},n_{ij}^{(2)},\ldots,n_{ij}^{(K)}]^T \in \mathbb{R}^{K \times 1}$.

For the sake of exposition, we consider the example of a network with $M=3$ anchors and one sensor ({\it node 0}) with a of two-way communication protocol between each of the sensor-anchor pairs. Let the clock parameters corresponding to the $i$th node be collected in a vector ${\bf c}_i = [\alpha_i,\beta_i]$ and $\boldsymbol{\tau}_0= [\tau_{0,1},\tau_{0,2},\ldots,\tau_{0,M}]^T \in\mathbb{R}^{M \times 1}$. The pairwise distances of the sensor to each anchor will be ${\bf d}_0 = \nu \boldsymbol{\tau}_0$. We can now write (\ref{eq:timemarkers}) for all the $K$ time-stamps collected in a matrix-vector form shown in (\ref{eq:example}) on top of this page.
Moving the known columns corresponding to ${\bf c}_3=[1,0]^T$ (clock reference) to one side, we can re-write (\ref{eq:example}) shown in (\ref{eq:example2}) on top of this page. 
The generalization of (\ref{eq:example2}) for any $M >2$ is straightforward.

In case we adopt the broadcast based ATPL protocol, the matrix $\bf A$ will have additional rows corresponding to the passive listening links~\cite{ChepuriSPL}. The detailed derivation of the linear data model for the ATPL protocol can be found in~\cite{ChepuriSPL}.

The generalized linear model for either the two-way communication or ATPL protocol can be written as  
\begin{equation}
\vspace*{-2mm}
\label{eq:datamodel}
\mathbf{A}{\boldsymbol{\theta}} = {\mathbf{t}} + \mathbf{n}
\end{equation}
where $\mathbf{A} \in \mathbb{R}^{KM \times 3M}$, $\boldsymbol{\theta} \in \mathbb{R}^{3M \times 1}$, ${\bf t} \in \mathbb{R}^{KM \times 1}$, and ${\bf n} \in \mathbb{R}^{KM \times 1}$ with the structures detailed in (\ref{eq:example2}) for the two-way communication protocol or in \cite{ChepuriSPL} for the ATPL protocol.

The aim of this work is to estimate the position ${\bf x}_0$ of the target node along with all the unknown clock parameters. 
The position of the target node ${\bf x}_0$ can be computed using the range estimates obtained by solving (\ref{eq:datamodel}). This is presented as a two-step approach in the next section, where in the first step we estimate the unknown clock parameters and the pairwise distances of the sensor to each anchor, and use this estimated pairwise distances to compute the position of the target node in the second step. Alternatively, we can formulate a single estimation problem to jointly compute the position of the target node as well as all the unknown clock parameters and this is the main contribution of this paper.

\section{Two-step approach}

In the two-step approach, we first solve for all the unknown clock parameters and the pairwise distances of the sensor to each anchor and then use the range estimates in a LS estimator to compute the location.
\vspace*{-4mm}
\subsection{Joint synchronization and ranging: step I}

For $K \geq 3$, matrix ${\bf A}$ is tall and left-invertible. Hence, the unknown parameters in $\boldsymbol{\theta}$ can be estimated using LS, i.e.,
\begin{equation}
\vspace*{-2mm}
\begin{aligned}
\label{eq:LSex}
{\hat{\boldsymbol{\theta}}} &=({\bf A}^{T}{\bf A})^{-1}{\bf A}^{T}\mathbf{t}.
\end{aligned}
\end{equation}
Subsequently, the clock-skews $\boldsymbol{\omega}$, clock-offsets $\boldsymbol{\phi}$ can be obtained using the relation in~(\ref{eq:omega}), and the pairwise distances of the sensor to each anchor using the relation $\hat{\mathbf{d}_0} = \nu\hat{\boldsymbol{\tau}_0}$. 
\vspace*{-4mm}
\subsection{Localization from estimated pairwise distances: step~II} \label{sec:twostage}
Pairwise distances form a major input to any localization scheme. Using the pairwise distance estimates obtained in (\ref{eq:LSex}), the coordinates of the sensor node can be estimated using range-squared localization algorithms.  

Let us define a vector $\mathbf{q} = [\|\mathbf{x}_{1}\|^2,\|\mathbf{x}_{2}\|^2,\ldots,\|\mathbf{x}_{{M}}\|^2]^T \in \mathbb{R}^{M \times 1}$. Using (\ref{eq:distance}), we can write the pairwise distance of the sensor to each anchor in a vector form as  
\begin{equation}\label{eq:distsq}
\begin{aligned}
{\mathbf{d}_0} \odot {\mathbf{d}_0} &= \mathbf{q} - 2\mathbf{X}^T\mathbf{x}_{0} + \|\mathbf{x}_{0}\|^2\mathbf{1}_M\\
&=   \bar{\bf{X}}\mathbf{p} + \mathbf{q},
\end{aligned}
\end{equation}
where $\mathbf{p} = [\mathbf{x}_{0}^T, \|\mathbf{x}_{0}\|^2]^T$ and $\bar{\mathbf{X}} = [- 2\mathbf{X}^T,\mathbf{1}_M] \in \mathbb{R}^{M \times (l+1)}$.
Subsequently, the coordinates of the sensor can be estimated using LS as follows
\begin{equation}
\vspace*{-1mm}
\begin{aligned}
\label{eq:LSposition}
{\hat{\bf p}} &=({\bar{\mathbf{X}}}^{T}{\bar{\mathbf{X}}})^{-1}{\bar{\mathbf{X}}}^{T}(\hat{\mathbf{d}_0} \odot \hat{\mathbf{d}_0} - {\bf q})
\end{aligned}
\end{equation} provided $M \geq l+1$ such that the matrix ${\bar{\mathbf{X}}}$ is tall. The anchor positions can be designed such that the matrix ${\bar{\mathbf{X}}}$ is left-invertible.
\vspace*{-5mm}
\section{The joint estimator}
Although clock synchronization and localization problems are tightly coupled, they have a non-linear relation as can be seen in (\ref{eq:distsq}). In case we want to localize the sensor 
and also estimate all the unknown clock parameters in one linear step, we have to linearize the relation between the clock parameters and the position. 

In order to do such a joint localization and synchronization, we exploit the linear relation between the range-squared measurements and the coordinates of the sensor. Instead of squaring the estimated pairwise distances in the second step as in Section \ref{sec:twostage}, we linearize the problem by squaring the linear model in (\ref{eq:datamodel}), and this is the main contribution of this paper. A unified framework for localization and synchronization is essential for applications in WSNs such as joint tracking of the position and the clock-parameters, for e.g., using a standard Kalman filter.

Hadamard squaring the data model in (\ref{eq:datamodel}) would result in a linear model and is given by
\begin{equation}
\vspace*{-1mm}
\begin{aligned}
\label{eq:Hadamard1}
(\mathbf{A}{\boldsymbol{\theta}}) \odot (\mathbf{A}{\boldsymbol{\theta}}) = (\mathbf{t} + \mathbf{n}) \odot (\mathbf{t} + \mathbf{n})
\end{aligned}
\end{equation} and can be further simplified to
\begin{equation}
\vspace*{-1mm}
\begin{aligned}
\label{eq:Hadamard1}
(\mathbf{A}^T \circ \mathbf{A}^T)^T ({\boldsymbol{\theta}} \otimes {\boldsymbol{\theta}}) = (\mathbf{t} + \mathbf{n}) \odot (\mathbf{t} + \mathbf{n})
\end{aligned}
\end{equation}
The linear model in (\ref{eq:Hadamard1}) does not have a unique solution as the system matrix $(\mathbf{A}^T \circ \mathbf{A}^T)^T$ is a $KM \times 9M^2$ matrix which is generally fat and thus not left-invertible. However, this can be solved with certain approximations of the clock parameters as in~\cite{yiyinATR}.

An alternative approach to linearizing the problem is by taking the Kroneckor product of the measurements, i.e., 
\begin{equation}
\vspace*{-1mm}
\begin{aligned}
\label{eq:propmodel}
(\mathbf{A}{\boldsymbol{\theta}}) \otimes (\mathbf{A}{\boldsymbol{\theta}}) = (\mathbf{t} + \mathbf{n}) \otimes (\mathbf{t} + \mathbf{n}). 
\end{aligned}
\end{equation}  Using the matrix property ${\bf P}{\bf C} \otimes {\bf Q}{\bf E} = ({\bf P} \otimes {\bf Q}) ({\bf C} \otimes {\bf E})$, we can further simplify  (\ref{eq:propmodel}) to the following linear model
\begin{equation}
\vspace*{-1mm}
\begin{aligned}
\label{eq:propmodel1}
\overbrace{({\bf A} \otimes {\bf A})}^{\bar{\mathbf{A}}} \overbrace{(\boldsymbol{\theta} \otimes \boldsymbol{\theta})}^{\bar{\boldsymbol{\theta}}} &= (\mathbf{t} + \mathbf{n}) \otimes (\mathbf{t} + \mathbf{n}) \\
&= \overbrace{{\bf t} \otimes {\bf t}}^{\bar{\bf t}} + \overbrace{{\bf n} \otimes {\bf n} + {\bf t} \otimes {\bf n} +  {\bf n} \otimes {\bf t}}^{\bf w}
\end{aligned}
\end{equation}  where $\bar{\bf A} \in \mathbb{R}^{K^2M^2 \times 9M^2}$, and ${\bf w} \in \mathbb{R}^{K^2M^2 \times 1}$ is the new error vector. If the matrix ${\bf A}$ is full column-rank, then it follows that the matrix $\bar{\bf A}$ is also full column-rank.

We now introduce two new variables to resolve the clock parameters without ambiguity after the squaring operation. For the $i$th node, we define the variables
\begin{eqnarray}
\gamma_i \triangleq \alpha_{i}^2 \quad \text{and} \quad \delta_i \triangleq \alpha_i\beta_{i} \label{eq:clckparam2},
\end{eqnarray}
and collect parameters corresponding to the $i$th node in the vector $\bar{\mathbf{c}}_{i}=[\gamma_{i}, \delta_{i}]^T$.  The clock-skew $\omega_i \in \mathbb{R}_+$ is always positive and the clock-offset $\phi_i \in \mathbb{R}$ can be either positive or negative.  As a result, recovering clock-offsets from $\beta_i^2$ without ambiguities would be difficult. Hence, we make use of the cross-term $\delta_i= \alpha_i\beta_i$ to recover the clock-offset. For all the nodes in the network we have $\bar{\mathbf{c}} = [\bar{\mathbf{c}}_{0}^T,\bar{\mathbf{c}}_{1}^T,\bar{\mathbf{c}}_{2}^T,\ldots,\bar{\mathbf{c}}_{M-1}^T]^T$. Let us define a permutation matrix $\boldsymbol{\Pi} \in \mathbb{R}^{9M^2 \times 9M^2 }$ that sorts the entries of  $\bar{\boldsymbol{\theta}}$, such that $\boldsymbol{\Pi}\bar{\boldsymbol{\theta}} = [\bar{\mathbf{c}}^T, \boldsymbol{\tau}_0^{\odot2T}, {\bf z}^T]^T$. Here, the entries of the vector ${\bf z} \in \mathbb{R}^{L_z \times 1}$ with $L_z = 9M^2-3M$, consist of the nuisance parameters excluding $\bar{\mathbf{c}}$ and $\boldsymbol{\tau}_0^{\odot2}$ from $\bar{\boldsymbol{\theta}}$ and is of less interest. 

We can now re-write (\ref{eq:propmodel1}) as follows
\begin{equation}
\vspace*{-1mm}
\begin{aligned}
\label{eq:propmodel2}
{\bar{\mathbf{A}}} \boldsymbol{\Pi}^T ({\bf S}_{\bar{\mathbf{c}}}\bar{\mathbf{c}} + \nu^{-2} {\bf S}_{{\mathbf{d}}} {\bf d}_0^{\odot2} +{\bf S}_{{\mathbf{z}}} {\bf z}) 
&= {\bar{\bf t}} + {\bf w}
\end{aligned}
\end{equation} where ${\bf S}_{\bar{\mathbf{c}}}$, ${\bf S}_{{\mathbf{d}}}$, and ${\bf S}_{\bar{\mathbf{z}}}$ are the selection matrices to select columns of ${\bar{\mathbf{A}}} \boldsymbol{\Pi}^T$ corresponding to $\bar{\mathbf{c}}$, ${\bf d}_0^{\odot2}$, and ${\bf z}$, respectively.
Substituting (\ref{eq:distsq}) in (\ref{eq:propmodel2}), we get 
\begin{equation}
\begin{aligned}
\small
\label{eq:propmodel3}
{\bar{\mathbf{A}}} \boldsymbol{\Pi}^T ({\bf S}_{\bar{\mathbf{c}}}\bar{\mathbf{c}} + \nu^{-2} {\bf S}_{{\mathbf{d}}}  \bar{\bf{X}}_a\mathbf{p} + {\bf S}_{{\mathbf{z}}} {\bf z}) 
&= {\bar{\bf t}}   -  {\bar{\mathbf{A}}} \boldsymbol{\Pi}^T \nu^{-2} {\bf S}_{{\mathbf{d}}} \mathbf{q} + {\bf w}. 
\end{aligned}
\end{equation}
We next collect the unknowns in the vector $\boldsymbol{\psi} = [\bar{\mathbf{c}}^T,\mathbf{p}^T,{\bf z}^T ]^T$ $\in \mathbb{R}^{L \times 1}$ where $L = 2M+l+1+L_z$ and the columns corresponding to the unknowns in the matrix 
\begin{equation}
\vspace*{-1mm}
\begin{aligned}
\label{eq:propmodel4}
\tilde{\bf A} = [{\bar{\mathbf{A}}} \boldsymbol{\Pi}^T{\bf S}_{\bar{\mathbf{c}}}, \nu^{-2}\bar{\mathbf{A}} \boldsymbol{\Pi}^T {\bf S}_{{\mathbf{d}}}\bar{\bf{X}} , {\bar{\mathbf{A}}} \boldsymbol{\Pi}^T{\bf S}_{{\mathbf{z}}}] \in \mathbb{R}^{K^2M^2 \times L},
\end{aligned}
\end{equation} and the measurements in the vector $\tilde{\bf t} = {\bar{\bf t}}   -  \nu^{-2} {\bar{\mathbf{A}}} \boldsymbol{\Pi}^T  {\bf S}_{{\mathbf{d}}} \mathbf{q} \in \mathbb{R}^{K^2M^2 \times 1}$. 

The generalized linear model for joint localization and synchronization is then given by
\begin{equation}
\vspace*{-1mm}
\begin{aligned}
\label{eq:propmodel5}
\tilde{\bf A} \boldsymbol{\psi} = \tilde{\bf t} + {\bf w}
\end{aligned}
\end{equation} The unknown parameters in $\boldsymbol{\varphi}$ can be estimated using LS, i.e., 
\begin{equation}
\vspace*{-1mm}
\begin{aligned}
\label{eq:propLS}
\hat{\boldsymbol{\psi}} = (\tilde{\bf A}^T \tilde{\bf A})^{-1}\tilde{\bf A} ^T\tilde{\bf t}.
\end{aligned}
\end{equation} Hence, the unknown position ${\bf x}_0$ is obtained by solving (\ref{eq:propLS}) and the unknown clock-skews and clock-offsets can be obtained using (\ref{eq:clckparam2}) and (\ref{eq:omega}) without any ambiguities.  

Alternatively, a weighted least-squares (WLS) estimator instead of (\ref{eq:LSposition}) taking the estimation error in (\ref{eq:LSex}) or a WLS estimator instead of (\ref{eq:propLS}) pre-whitening the noise ${\bf w}$ is possible. However, this is not further detailed in this paper.

\section{Cram\'er-Rao lower bound}
We now derive the CRLB for jointly estimating the clock-skews $\boldsymbol{\omega}$, the clock-offsets $\boldsymbol{\phi}$, and the coordinates of the sensor node $\mathbf{x}_0$, i.e., $\bar{\boldsymbol{\psi}}=[\boldsymbol{\omega}^T,\boldsymbol{\phi}^T,\mathbf{x}_0^T]^T$ based on (\ref{eq:datamodel}).  For an unbiased estimator $\hat{\bar{\boldsymbol{\varphi}}}$ it follows from the CRLB theorem that $\mathbb{E}(\hat{\bar{\boldsymbol{\psi}}}\hat{\bar{\boldsymbol{\psi}}}^T) \geq {\bf F}^{-1}$
where $\mathbf{F}$ is the Fisher information matrix.
If the error vector $\mathbf{n}$ is Gaussian distributed with a variance $\sigma^2$, then $\mathbf{F}$ can be computed as $\mathbf{F} = \sigma^{-2} \mathbf{J}^T\mathbf{J}$, 
where $\mathbf{J}$ is the Jacobian matrix given by
\begin{equation}\label{eq:FIM}
\begin{aligned}
{\mathbf{J}} &= \frac{\partial(\mathbf{A}\boldsymbol{\theta}-\mathbf{t})}{\partial\bar{\boldsymbol{\varphi}}} = [{\mathbf{J}}_{\boldsymbol{\omega}} \quad {\mathbf{J}}_{\boldsymbol{\phi}} \quad {\mathbf{J}}_{\mathbf{x}_0}] \in \mathbb{R}^{KM \times (2M+l)}
\end{aligned}
\end{equation}
with sub-blocks 
\begin{equation}
\label{eq:Jsubmatrix}
\begin{aligned}
{\mathbf{J}}_{\boldsymbol{\omega}} &= - \mathbf{A}({\mathbf{S}}_{\boldsymbol{\alpha}} -{\mathbf{S}}_{\boldsymbol{\beta}} \odot \mathbf{1}_{KM}\boldsymbol{\phi}^T) \oslash (\mathbf{1}_{KM}\boldsymbol{\omega}^T)^{\odot2}, \\
{\mathbf{J}}_{\boldsymbol{\phi}} &= - \mathbf{A}{\mathbf{S}}_{\boldsymbol{\beta}}  \oslash \mathbf{1}_{KM}\boldsymbol{\omega}^T,\\
{\mathbf{J}}_{\mathbf{x}_0} &= \nu^{-1}\mathbf{T}{\mathbf{S}}_{\boldsymbol{\tau}_0}{\bf D},
\end{aligned}
\end{equation}
where 
 ${\mathbf{S}}_{\boldsymbol{\alpha}}$, ${\mathbf{S}}_{\boldsymbol{\beta}}$, and ${\mathbf{S}}_{\boldsymbol{\tau}_0}$ are selection matrices to select the columns of ${\bf A}$ corresponding to $\boldsymbol{\alpha}$, $\boldsymbol{\beta}$, and $\boldsymbol{\tau}_0$, respectively. The $M \times l$ derivative matrix ${\bf D}$ is defined as
\begin{eqnarray}
&& {[{\bf D}]}_{i,j}={\left[\frac{\partial\mathbf{d}_0}{\partial\mathbf{x}_0}\right]}_{i,j} = \frac{[{{\bf x}_{0}]}_j- [{{\bf x}_{i}]}_j}{{\|\mathbf{x}_0-\mathbf{x}_{i}\|}_2}
\end{eqnarray} 

\section{Numerical example}
A network with one target sensor and $5$ anchors is considered.  Both the target node and anchor nodes are deployed randomly within a range of $100\mathrm{m}$. 
The clock-skews $\boldsymbol{\omega}$ and clock-offsets $\boldsymbol{\phi}$ are uniformly distributed in the range $[1-100\mathrm{ppm},1+100\mathrm{ppm}]$ and $[-1\mathrm{s},1\mathrm{s}]$, respectively. We use an observation interval of $100 \,\mathrm{s}$ during which the clock parameters are assumed to be fixed. We use $\nu = 300 \, \mathrm{m/s}$ and record $K=10$ time-stamps. The error vector is assumed to be Gaussian distributed with a variance $\sigma^2$. The simulations are averaged over $1000$ independent Monte Carlo experiments. 
\begin{figure} [!h]
\centering
\includegraphics[height=2in,width=\columnwidth]{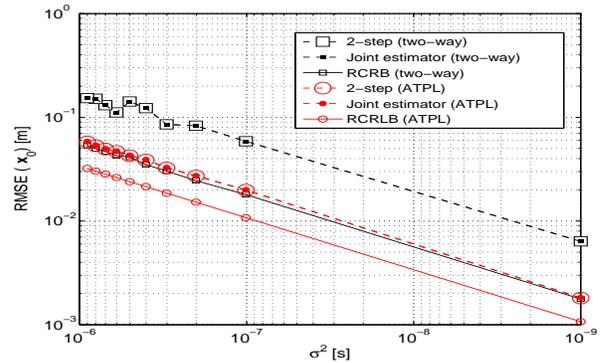}
\caption{RMSE of the estimated sensor coordinates.}
\label{fig:RMSEx}
\end{figure} 
\begin{figure} [!h]
\centering
\includegraphics[height=2in,width=\columnwidth]{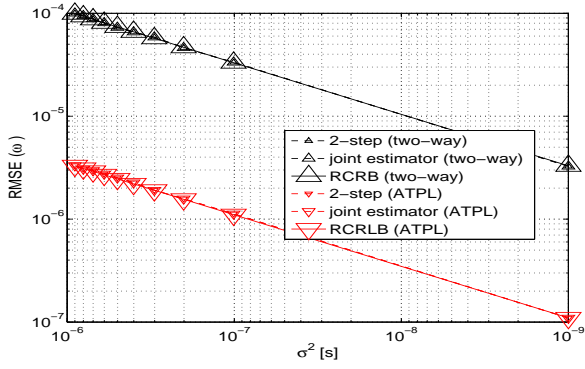}
\caption{RMSE of the estimated clock-skews.}
\label{fig:RMSEomega}
\end{figure} 
\begin{figure} [!h]
\centering
\includegraphics[height=2in,width=\columnwidth]{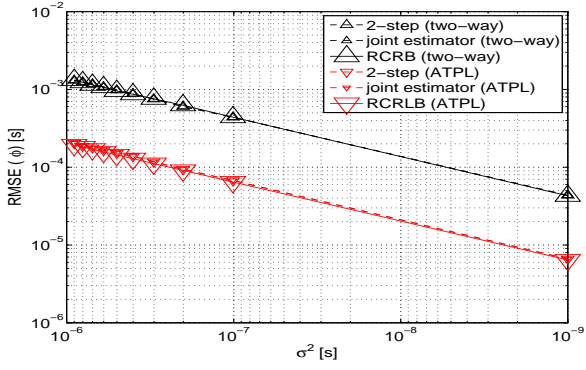}
\caption{RMSE of the estimated clock-offsets.}
\label{fig:RMSEphi}
\end{figure} 

In this paper, we analyze the performance of the proposed estimators in terms of the root mean square error (RMSE) of the estimated sensor position, clock-skews, and clock-offsets for different values of $\sigma^2$. We provide the results for a) two-way communication protocol~\cite{raj_camsap} b) ATPL protocol~\cite{ChepuriSPL}.

Fig. \ref{fig:RMSEx} shows the RMSE of the estimated sensor position computed using a) the two-step approach, i.e., LS estimator (\ref{eq:LSposition}) with the range estimates of (\ref{eq:LSex}) as the input, and b) the proposed joint estimator. The root CRLB (RCRLB) is also provided for both the considered protocols. The ATPL protocol performs better than the two-way communication protocol due to the additional passive listening links~\cite{ChepuriSPL}. However, both the estimators for localization are inaccurate as the dependencies with the nuisance parameters are not considered, and this can be resolved using a constrained WLS solutions~\cite{yiyinATR}. 

Fig. \ref{fig:RMSEomega} and Fig. \ref{fig:RMSEphi} show the RMSE of the estimated clock-skews and clock-offsets.  The ATPL protocol again performs better than the two-way communication protocol. In addition, both the estimators for clock-skews and clock-offsets are asymptotically efficient, and meet the CRLB.

The location of the target node can be obtained with a two-step approach using the range estimates obtained from joint synchronization and ranging. Alternatively, we can formulate  localization and synchronization under a unified framework as a single linear problem. However, this results in a larger system to solve and is computationally less attractive than the two-step approach. The linear model of the joint synchronization and localization can be used for joint tracking of the clock parameters and the position which is an important application in a WSN. 

\section{Conclusions}
We have considered a fully-asynchronous network with one sensor and a few anchors. In this paper, we have addressed a problem in which we estimate all the unknown clock parameters as well as the position of the target node. Location of the node can be estimated with a two-step approach using the range estimates. To avoid this two-step approach, we have proposed a generic linear data model for joint localization and clock synchronization. An estimator based on LS to jointly estimate the position of the target sensor node, along with all the unknown clock-skews and clock-offsets has been presented. The proposed estimator for clock-skews and clock-offsets is asymptotically efficient and meets the CRLB, however, the position estimates do not asymptotically achieve the CRLB.
\bibliographystyle{IEEEbib}
\bibliography{IEEEabrv,strings,refs}
\end{document}